\documentclass[conference]{IEEEtran}

\usepackage[pdftex]{graphicx}
\usepackage{xcolor}
\usepackage[hidelinks]{hyperref}

\usepackage{balance}

\definecolor{Paired-1}{RGB}{31,120,180}
\definecolor{Paired-2}{RGB}{166,206,227}
\definecolor{Paired-3}{RGB}{51,160,44}
\definecolor{Paired-4}{RGB}{178,223,138}
\definecolor{Paired-5}{RGB}{227,26,28}
\definecolor{Paired-14}{RGB}{251,154,153}
\definecolor{Paired-7}{RGB}{255,127,0}
\definecolor{Paired-8}{RGB}{253,191,111}
\definecolor{Paired-9}{RGB}{106,61,154}
\definecolor{Paired-10}{RGB}{202,178,214}
\definecolor{Paired-11}{RGB}{177,89,40}
\definecolor{Paired-12}{RGB}{255,255,153}
\definecolor{Paired-13}{RGB}{80,80,80}
\definecolor{Paired-14}{RGB}{153, 153, 255}

\usepackage{amsmath,amssymb,amsfonts,stmaryrd}
\usepackage{bm}
\usepackage{dsfont}
\usepackage{numprint}

\usepackage{adjustbox}
\usepackage{booktabs}
\usepackage{multirow}

\usepackage[caption=false,font=footnotesize]{subfig}
\usepackage{tikz}
\usetikzlibrary{matrix, positioning, patterns, shapes, arrows}
\usepackage{pgfplots}
\usepackage{amsmath}
\newcommand{\argmin}{\mathop{\mathrm{argmin}}\limits} 
\usepgfplotslibrary{groupplots}
\usepgfplotslibrary{colorbrewer}
\usetikzlibrary{spy,backgrounds}

\usepackage{pgfplots}
\pgfplotsset{compat=newest}
\usepgfplotslibrary{groupplots}

\usepackage{tikz}
\usetikzlibrary{matrix, positioning, patterns, shapes, arrows}

\usepackage[draft]{fixme} 
\fxsetup{theme=color}

\usepackage{lipsum}

\usepackage[ruled, linesnumbered, vlined]{algorithm2e}

\title{GRAND for Rayleigh Fading Channels}

\author{\IEEEauthorblockN{Syed Mohsin Abbas, Marwan Jalaleddine and Warren J. Gross}
\IEEEauthorblockA{Department of Electrical and Computer Engineering\\
McGill University,  Montr\'eal, Qu\'ebec, Canada\\
Emails: syed.abbas@mail.mcgill.ca, marwan.jalaleddine@mail.mcgill.ca, warren.gross@mcgill.ca}}

\begin{document}

\maketitle

\begin{abstract}
Guessing Random Additive Noise Decoding (GRAND) is a code-agnostic decoding technique for short-length and high-rate channel codes. GRAND attempts to guess the channel-induced noise by generating Test Error Patterns (TEPs), and the sequence of TEP generation is the primary distinction between GRAND variants. In this work, we extend the application of GRAND to multipath frequency non-selective Rayleigh fading communication channels, and we refer to this GRAND variant as Fading-GRAND. The proposed Fading-GRAND adapts its TEP generation to the fading conditions of the underlying communication channel, outperforming traditional channel code decoders in scenarios with $L$ spatial diversity branches as well as scenarios with no diversity. Numerical simulation results show that the Fading-GRAND outperforms the traditional Berlekamp-Massey (B-M) decoder for decoding BCH code $(127,106)$ and BCH code $(127,113)$ by $\mathbf{0.5\sim6.5}$ dB at a target FER of $10^{-7}$. Similarly, Fading-GRAND outperforms GRANDAB, the hard-input variation of GRAND, by $0.2\sim8$ dB at a target FER of $10^{-7}$ with CRC $(128,104)$ code and RLC $(128,104)$. Furthermore the average complexity of Fading-GRAND, at $\frac{E_b}{N_0}$ corresponding to target FER of $10^{-7}$, is $\frac{1}{2}\times\sim \frac{1}{46}\times$ the complexity of GRANDAB.

\end{abstract}


\section{Introduction}

GRAND is a universal Maximum likelihood decoding technique for high-rate and short-length channel codes \cite{Duffy19TIT,valembois}. These short codes are appealing for ultra-reliable low-latency communication (URLLC) \cite{URLLC2}, the internet of things (IoT) \cite{IoT1,IoT2}, massive machine type communication (mMTC) \cite{MTC_1}, wireless sensor networks (WSN) \cite{MTC_2}, as well as many other novel emerging applications. 
GRAND guesses the channel induced noise by generating the TEPs and combining them with the received vector to reverse the effect of the noise. GRAND features hard-input and soft-input variants that differ primarily in how the TEPs are generated. 

GRAND's universal decoding capability opens up new possibilities, and it has been demonstrated to work with memoryless communication channels as well as  channels with memory \cite{GRANDMO}. GRAND adjusts the generation of TEPs in response to channel conditions. The GRAND Markov Order (GRAND-MO) \cite{GRANDMO}, for example, uses noise correlations and adapts its TEP generation to mitigate the effect of noise bursts in communication channels with memory.  

In dense urban settings, channel fading can cause significant performance degradation of a wireless communication system. Traditionally, diversity techniques (time, frequency, and space) are used to mitigate multipath fading and improve performance of a wireless communications system  \cite{tse2005fundamentals}. In this paper, we propose Fading-GRAND, a novel hard-input variant of GRAND that adapts the TEP generation to the fading conditions of the channel. Fading-GRAND outperforms traditional code decoders over the multipath flat Rayleigh fading communication channel; a channel that models the effect of small-scale fading in a multipath propagation environment with no dominant line of sight between the transmitter and the receiver \cite{Rayleighbook}. We also investigate the proposed Fading-GRAND with SIMO (single input, multiple output) spatial diversity, which corresponds to a scenario in which the transmitter has a single transmit antenna and the receiver has multiple antennas.

In the absence of any spatial diversity, the Fading-GRAND outperforms the Berlekamp-Massey (B-M) decoder \cite{Berlekamp68,Massey69} in decoding BCH \cite{Hocquenghem59,Bose1960} code $(127,106)$ and BCH Code $(127,113)$ by $4\sim6.5$dB at a target FER of $10^{-7}$. Furthermore, at a target FER of $10^{-7}$, the Fading-GRAND outperforms the B-M decoder by $0.5\sim3$dB when the number of spatial diversity branches is increased. Similarly, at a target FER of $10^{-7}$, the Fading-GRAND outperforms the predecessor hard-input GRANDAB by $0.2\sim8$dB when decoding CRC $(128,104)$ code and RLC $(128,104)$.

Furthermore the average complexity of Fading-GRAND which corresponds to the average number of TEPs required, at $\frac{E_b}{N_0}$ corresponding to target FER of $10^{-7}$, is $\frac{1}{2}\times\sim \frac{1}{46}\times$ the complexity of GRANDAB. It should be noted that the low average complexity of GRAND and its variants directly translates to low decoding latency as well as low energy consumption, which is demonstrated by the high-throughput and energy-efficient GRAND hardware architecture \cite{GRANDAB-VLSI} rendering GRAND and its variants appealing for applications with stringent latency requirements and on a tight energy consumption budget \cite{IoT1}-\cite{MTC_2}.

The rest of this paper is organized as follows: The channel model under consideration is explained in Section II. The proposed Fading-GRAND is presented in Section III, while the numerical simulation results are presented in Section IV. Finally, in Section V, concluding remarks are made.

\section{Preliminaries}
\subsection{Notations}
Matrices are denoted by a bold upper-case letter ($\bm{M}$), while vectors are denoted with bold lower-case letters ($\bm{v}$). The transpose operator is represented by $^\top$. The $i^{\text{th}}$ element of a vector $\bm{v}$ is denoted by $v_i$. The number of $k$-combinations from a given set of $n$ elements is noted by $\binom{n}{k}$. $\mathds{1}_n$ is the indicator vector where all locations except the $n^{\text{th}}$ location are $0$ and the $n^{\text{th}}$ location is $1$. Similarly, $\mathds{1}_{i,j\ldots,z}=\mathds{1}_i\oplus\mathds{1}_j\oplus\ldots\mathds{1}_z$, with $i\neq j \ldots\neq k$.
All the indices start at $1$. For this work, all operations are restricted to the Galois field with 2 elements, noted $\mathbb{F}_2$. Furthermore, we restrict ourselves to $(n,k)$ linear block codes, where $n$ is the code length and $k$ is the code dimension. The symbol $\iff$ denotes \textit{if and only if}. The conjugate of a complex number $h$ is denoted by $h^{\ast}$.

\subsection{Channel Model}

In this paper, we consider $\sqrt{E_{b}}c$ to represent the transmitted binary phase-shift keying (BPSK) signal, where $c=\pm1$ with equal probability. The signal is transmitted by one transmit antenna over a slow frequency-nonselective Rayleigh fading channel. We also assume that there are $L$ receive antennas which are sufficiently spaced apart. Hence, the gains $h_i~(\forall i \in [1, L])$ are independent Rayleigh distributed, and we get a diversity gain of $L$ \cite{tse2005fundamentals}. The low-pass equivalent received signal at the $i^{th}$ antenna is given by (\ref{eq:chn_1}) \cite{ChannelModel_1}.

\begin{equation}
\hspace{1cm} y_i=h_{i}\sqrt{E_{b}}c+n_{i} \hspace{1cm}i=1\cdots L
\label{eq:chn_1}
\end{equation}

The noise $n_{i}~(\forall i \in [1, L])$ is a complex-valued $(n_{i}\in\mathbb{C})$ additive white Gaussian noise (AWGN) with a variance $\frac{N_0}{2}$ per complex dimension. $h_{i}$ is the multipath fading gain, which is complex valued and has in-phase (real) and quadrature (imaginary) components that are assumed to be Gaussian, stationary, and independent (orthogonal) of each other. The fading channel amplitude $\vert h_{i}\vert \geq 0$ follows a Rayleigh probability distribution, and the average power gain is normalized to unity ($\mathbb{E}[\vert h_{i}\vert^2]=1$). In this work, we assume that the receiver has perfect knowledge of the channel state information (CSI) \cite{benedetto1999principles}. Furthermore, because of perfect interleaving (i.e., infinite-depth), the $h_{i}$ affecting the various symbols are independent of one another \cite{tse2005fundamentals}. 

To combine $L$ independent spatial diversity branches, we consider selection combining (SC) and maximal ratio combining (MRC), followed by demodulation. SC is the simplest diversity combining technique, in which the diversity branch ($j$) with the largest instantaneous signal power among $L$ available diversity branches is selected for signal detection \cite{ChannelModel_1}.
\begin{equation}
y=h_{\max}^{\ast }(h_{\max}\sqrt{E_{b}}c+n_j),  ~~h_{\max} = max(|h_1|, ... |h_L|)
\label{eq:chn_2}
\end{equation}

 MRC, on the other hand, is the optimal combining scheme that achieves maximum array gain by coherently combining all $L$ diversity branches \cite{ChannelModel_2}\cite{benedetto1999principles}.
\begin{equation}
y=\sum_{i=1}^{L}\frac{{h_{i}}^{*}}{\sqrt{\sum_{z=1}^{L}\vert h_{z}\vert}}(h_{i}\sqrt{E_{b}}c+n_{i})~~
\label{eq:chn_3}
\end{equation}

\subsection{GRAND Decoding}

For a $(n,k)$ linear block code with codebook $\mathcal{C}$, a vector $\bm{u}$ of size $k$ maps to a vector $\bm{c}$ of size $n$, and the ratio  $R \triangleq \frac{k}{n}$ is known as the code-rate. Furthermore, there exists a $k \times n$ matrix $\bm{G}$ called generator matrix ($\bm{c}\triangleq\bm{u}\cdot\bm{G}$) and a $(n-k)\times n$ matrix $\bm{H}$ called parity check matrix such that 

\begin{equation}
\hspace{1cm} ~\bm{H}\cdot{\bm{c}}^\top = \bm{0} \hspace{1cm} \forall~\bm{c} \in \mathcal{C}
\label{eq:constraint}
\end{equation}

GRAND guesses the channel noise that corrupted the transmitted codeword ($\bm{c}$). To that end, GRAND first starts by generating the TEPs ($\bm{e}$), then combines them with the hard-decided received vector ($\hat{\bm{y}}$) of channel observation values (demodulated symbols) and finally evaluates if the resulting vector ($\hat{\bm{y}} \oplus \bm{e}$) is a member of the codebook $\mathcal{C}$ ( \ref{eq:constraint}). If this is the case, the decoding is assumed to be successful, and $\bm{e}$ is declared as the guessed noise, while $\hat{\bm{c}} \triangleq \hat{\bm{y}}~\oplus~\bm{e}$ is outputted as the estimated codeword.
\begin{algorithm}[!t]
\caption{\label{alg:Fgrand}Fading GRAND}
    \DontPrintSemicolon
    \SetAlgoVlined  
    \SetKwData{e}{$\bm{e}$}
    \SetKwData{d}{$\lfloor\frac{d}{2}\rfloor$}
    \SetKwData{b}{${b}$}
    \SetKwData{g}{${g}$}
    \SetKwData{ABANDON}{${ABANDON}$}
    \SetKwData{estm}{$\hat{\bm{u}}$}
    \SetKwData{ginv}{$\bm{G}^{-1}$}
    \SetKwData{yhat}{$\hat{\bm{y}}$}
    \SetKwData{ind}{$\mathbb{I}$}
    \SetKwData{HW}{$AB$}
    \KwIn{\yhat, $\bm{H}$, \ginv, \ind, \HW}
    \KwOut{\estm}
    $\e \leftarrow \bm{0}$\;
    \If{$\bm{H} \cdot\yhat^\top == \bm{0}$}{
    $\estm \leftarrow \yhat\cdot\ginv$\;
            \KwRet{\estm}
    }
    \Else {
    \For{$HW \gets 1$ to \HW}{   
      $\e \leftarrow \mathds{1}_{\{\lambda_1, \lambda_2, \ldots, \lambda_{HW}\}}~\forall\{\lambda_1, \lambda_2, \ldots, \lambda_{HW}\}\not\in \ind$ \tcp*[r]{$\lambda_i\in[1,n],~i\in[1,HW]$} 
    \If{$\bm{H} \cdot(\yhat \oplus \e)^\top == \bm{0}$} {
            $\estm \leftarrow (\yhat \oplus \e)\cdot\ginv$\;
            \KwRet{\estm}
            }
    }
    }
\vspace*{-0.2em} 
\end{algorithm}
\begin{figure}[!tb]
  \centering
  \includegraphics[width=1\linewidth]{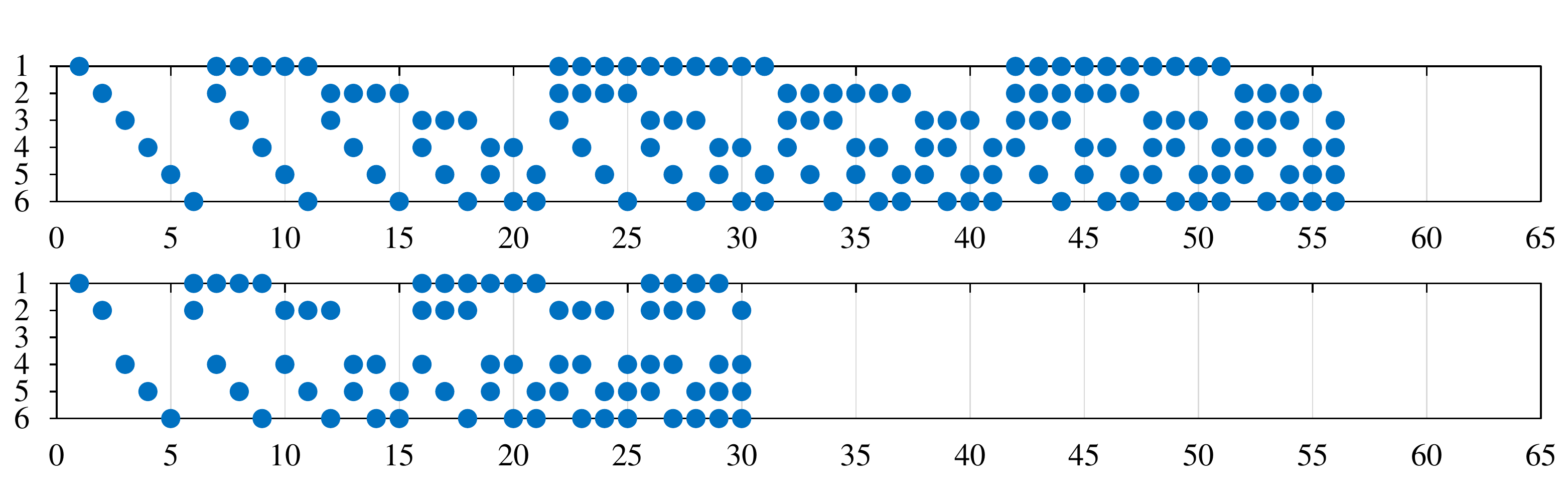}
  \vspace*{-2em}
  \caption{TEP generation for $n=6$. (a) Upper: For GRANDAB ($AB=4$) (b) Bottom: For Fading-GRAND ($AB=4$, $\mathbb{I}=\{3\}$)}
  \label{fig:TEP_GRANDAB}
  \vspace*{-1em} 
\end{figure}
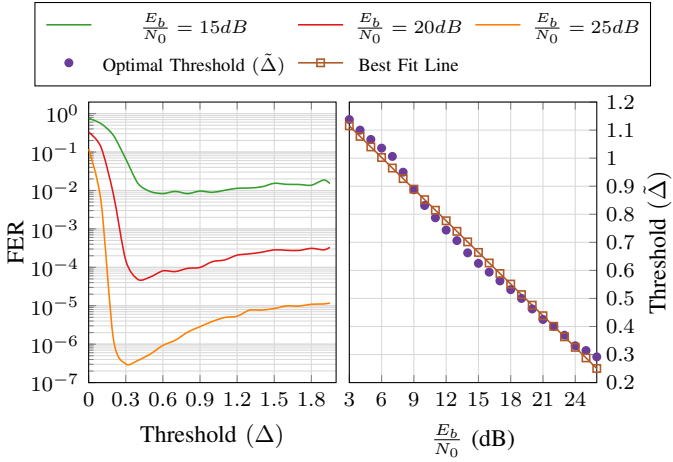
\begin{figure}[!tb]
  \centering
  \begin{tikzpicture}[]
    \begin{groupplot}[group style={group name=fer_queries, group size= 2 by 1, horizontal sep=5pt, vertical sep=5pt},
      footnotesize,
      height=.6\columnwidth,  width=0.55\columnwidth,
      tick align=inside,
      grid=both, grid style={gray!30},
      /pgfplots/table/ignore chars={|},
      ]

      \nextgroupplot[xlabel=Threshold $(\Delta)$, ylabel=FER, xmin=0, xmax=2, xtick={0,0.3,...,2.1},ytick pos=left, y label style={at={(axis description cs:-0.225,.5)},anchor=south},ymin=1.0e-7,  ymax=2, ymode=log]

      \addplot[smooth, Paired-3, semithick]  table[x=Threshold, y=FER] {data_fading/RLC/128_104/FER_FadingGrand_findParameter_15dB.txt};\label{gp:param_15}
       \addplot[smooth, Paired-5, semithick]  table[x=Threshold, y=FER] {data_fading/RLC/128_104/FER_FadingGrand_findParameter_20dB.txt};\label{gp:param_20}
       \addplot[smooth, Paired-7, semithick]  table[x=Threshold, y=FER] {data_fading/RLC/128_104/FER_FadingGrand_findParameter_25dB.txt};\label{gp:param_25}

      \coordinate (top) at (rel axis cs:0,1);

      \nextgroupplot[xlabel=$\frac{E_b}{N_0}$ (dB), ylabel=Threshold $(\tilde{\Delta})$, ytick pos=right,y label style={at={(axis description cs:1.340,.5)},anchor=south},xmin=3, xmax=26, xtick={3,6,...,27}, ymin=0.2, ymax = 1.2,]
      \addplot[only marks,mark options={scale=0.7}, Paired-9, semithick]  table[x=SNR, y=Threshold] {data_fading/RLC/128_104/RLC_128_104_Line.txt};\label{gp:line_1}
       \addplot[mark=square,mark options={scale=0.7} , Paired-11, semithick]  table[x=SNR, y=Line] {data_fading/RLC/128_104/RLC_128_104_Line.txt};\label{gp:line_2}

      \coordinate (bot) at (rel axis cs:1,0);
    \end{groupplot}
    \path (top|-current bounding box.north) -- coordinate(legendpos) (bot|-current bounding box.north);
    \matrix[
    matrix of nodes,
    anchor=south,
    draw,
    inner sep=0.2em,
    draw
    ]at(legendpos)
    {
      \ref{gp:param_15}& \scriptsize $\frac{E_b}{N_0}=15dB$  &[0.5pt]
      \ref{gp:param_20}& \scriptsize $\frac{E_b}{N_0}=20dB$  &[0.5pt]
      \ref{gp:param_25}& \scriptsize $\frac{E_b}{N_0}=25dB$ \\
      \ref{gp:line_1}& \scriptsize Optimal Threshold $(\tilde{\Delta})$  &[0.5pt]
      \ref{gp:line_2}& \scriptsize Best Fit Line \\
       }; 
  \end{tikzpicture}
  \vspace*{-2em}
  \caption{\label{fig:line_RLC_128_104} (a) (Left) Finding the optimal threshold $\tilde{\Delta}$ for RLC $(128,104)$ Fading-GRAND decoding. (b) (Right) Best Fit Line ($\tilde{\Delta}=m\times\frac{E_b}{N_0} + b$; $m= -0.0376$, $b = 1.228$).}
 \vspace*{-2em}
\end{figure}
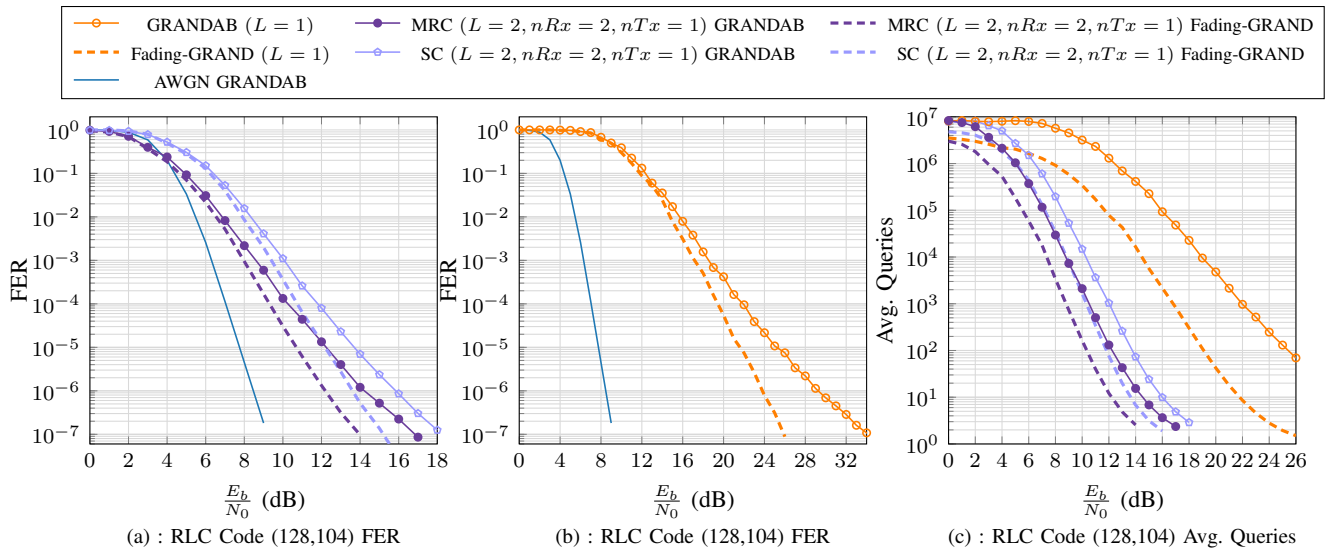
\begin{figure*}[ht]
  \centering
  \begin{tikzpicture}[spy using outlines = {rectangle, magnification=2.0, connect spies}]
    \begin{groupplot}[group style={group name=fer_queries, group size= 3 by 1, horizontal sep=31pt, vertical sep=45pt},
      footnotesize,
      width=.7\columnwidth, height=.67\columnwidth, 
      xlabel=$\frac{E_b}{N_0}$ (dB),
      xmin=0, xmax=36, xtick={0,2,...,34},
      ymode=log,
      tick align=inside,
      grid=both, grid style={gray!30},
      /pgfplots/table/ignore chars={|},
      ]

\nextgroupplot[ylabel= FER, ytick pos=left, y label style={at={(axis description cs:-0.155,.5)},anchor=south},ymin=6e-8, ymax = 2,xmin=0, xmax=18]
      \addplot[mark=none , Paired-1, semithick]  table[x=SNR, y=FER] {data_fading/RLC/128_104/FER_AWGN.txt};\label{gp:plot_RLC_a}

      \addplot[mark=*,mark options={scale=0.7} , Paired-9, semithick]  table[x=SNR, y=FER] {data_fading/RLC/128_104/FER_Fading_MRC_Div_2.txt};\label{gp:plot_RLC_e}
      \addplot[mark=pentagon,mark options={scale=0.7} , Paired-14, semithick]  table[x=SNR, y=FER] {data_fading/RLC/128_104/FER_Fading_SelComb_Div_2.txt};\label{gp:plot_RLC_g}
      \addplot[mark=none , Paired-9, densely dashed, very thick]  table[x=SNR, y=FER] {data_fading/RLC/128_104/FER_Fading_MRC_SRGrand_bestParameter.txt};\label{gp:plot_RLC_j}
      \addplot[mark=none , Paired-14, densely dashed, very thick]  table[x=SNR, y=FER] {data_fading/RLC/128_104/FER_Fading_Sel_SRGrand_bestParameter.txt};\label{gp:plot_RLC_l}

      \coordinate (top) at (rel axis cs:0,1);

      \nextgroupplot[ylabel= FER, ytick pos=left, y label style={at={(axis description cs:-0.155,.5)},anchor=south},ymin=6e-8, ymax = 2, xmin=0, xmax=34, xtick={0,4,...,36},]
      \addplot[mark=none , Paired-1, semithick]  table[x=SNR, y=FER] {data_fading/RLC/128_104/FER_AWGN.txt};
      
      \addplot[mark=o,mark options={scale=0.7} , Paired-7, semithick]  table[x=SNR, y=FER] {data_fading/RLC/128_104/FER_Fading.txt};\label{gp:plot_RLC_d}
      \addplot[mark=none , Paired-7, densely dashed, very thick]  table[x=SNR, y=FER] {data_fading/RLC/128_104/FER_Fading_SRGrand_bestParameter.txt};\label{gp:plot_RLC_i}

      \nextgroupplot[ylabel=Avg. Queries, ytick pos=left,y label style={at={(axis description cs:-0.12,.5)},anchor=south},ymin=1, ymax = 1e7, xmin=0, xmax=26,xtick={0,2,...,26},]

      \addplot[mark=pentagon,mark options={scale=0.7} , Paired-14, semithick]  table[x=SNR, y=Comp] {data_fading/RLC/128_104/FER_Fading_SelComb_Div_2.txt};
      \addplot[mark=none , Paired-14, densely dashed, very thick]  table[x=SNR, y=Comp] {data_fading/RLC/128_104/FER_Fading_Sel_SRGrand_bestParameter.txt};

      \addplot[mark=o,mark options={scale=0.7} , Paired-7, semithick]  table[x=SNR, y=Comp] {data_fading/RLC/128_104/FER_Fading.txt};
      \addplot[mark=none , Paired-7, densely dashed, very thick]  table[x=SNR, y=Comp] {data_fading/RLC/128_104/FER_Fading_SRGrand_bestParameter.txt};

      \addplot[mark=*,mark options={scale=0.7} , Paired-9, semithick]  table[x=SNR, y=Comp] {data_fading/RLC/128_104/FER_Fading_MRC_Div_2.txt};
      \addplot[mark=none , Paired-9, densely dashed, very thick]  table[x=SNR, y=Comp] {data_fading/RLC/128_104/FER_Fading_MRC_SRGrand_bestParameter.txt};


      \coordinate (bot) at (rel axis cs:1,0);
    \end{groupplot}
    \node[below = 1cm of fer_queries c1r1.south] {\footnotesize (a) : RLC Code (128,104) FER};
    \node[below = 1cm of fer_queries c2r1.south] {\footnotesize (b) : RLC Code (128,104) FER};
    \node[below = 1cm of fer_queries c3r1.south] {\footnotesize (c) : RLC Code (128,104) Avg. Queries};
    \path (top|-current bounding box.north) -- coordinate(legendpos) (bot|-current bounding box.north);
    \matrix[
    matrix of nodes,
    anchor=south,
    draw,
    inner sep=0.2em,
    draw
    ]at(legendpos)
    { \ref{gp:plot_RLC_d}& \scriptsize  GRANDAB $(L=1)$  &[5pt]
      \ref{gp:plot_RLC_e}& \scriptsize MRC $(L = 2, nRx=2, nTx=1)$  GRANDAB  &[5pt]
      \ref{gp:plot_RLC_j}& \scriptsize MRC $(L = 2, nRx=2, nTx=1)$ Fading-GRAND \\
      \ref{gp:plot_RLC_i}& \scriptsize  Fading-GRAND $(L=1)$ &[5pt]
      \ref{gp:plot_RLC_g}& \scriptsize SC $(L = 2, nRx=2, nTx=1) $ GRANDAB  &[5pt]
      \ref{gp:plot_RLC_l}& \scriptsize SC $(L = 2, nRx=2, nTx=1)$ Fading-GRAND \\
      \ref{gp:plot_RLC_a}& \scriptsize AWGN GRANDAB   \\
       }; 

  \end{tikzpicture}
  \vspace*{-1em}
  \caption{\label{fig:fer_RLC_128_104} Comparison of decoding performance and average complexity of RLC $(128,104)$ decoding via Fading-GRAND and GRANDAB decoder. }
  \vspace*{-1em} 
\end{figure*}


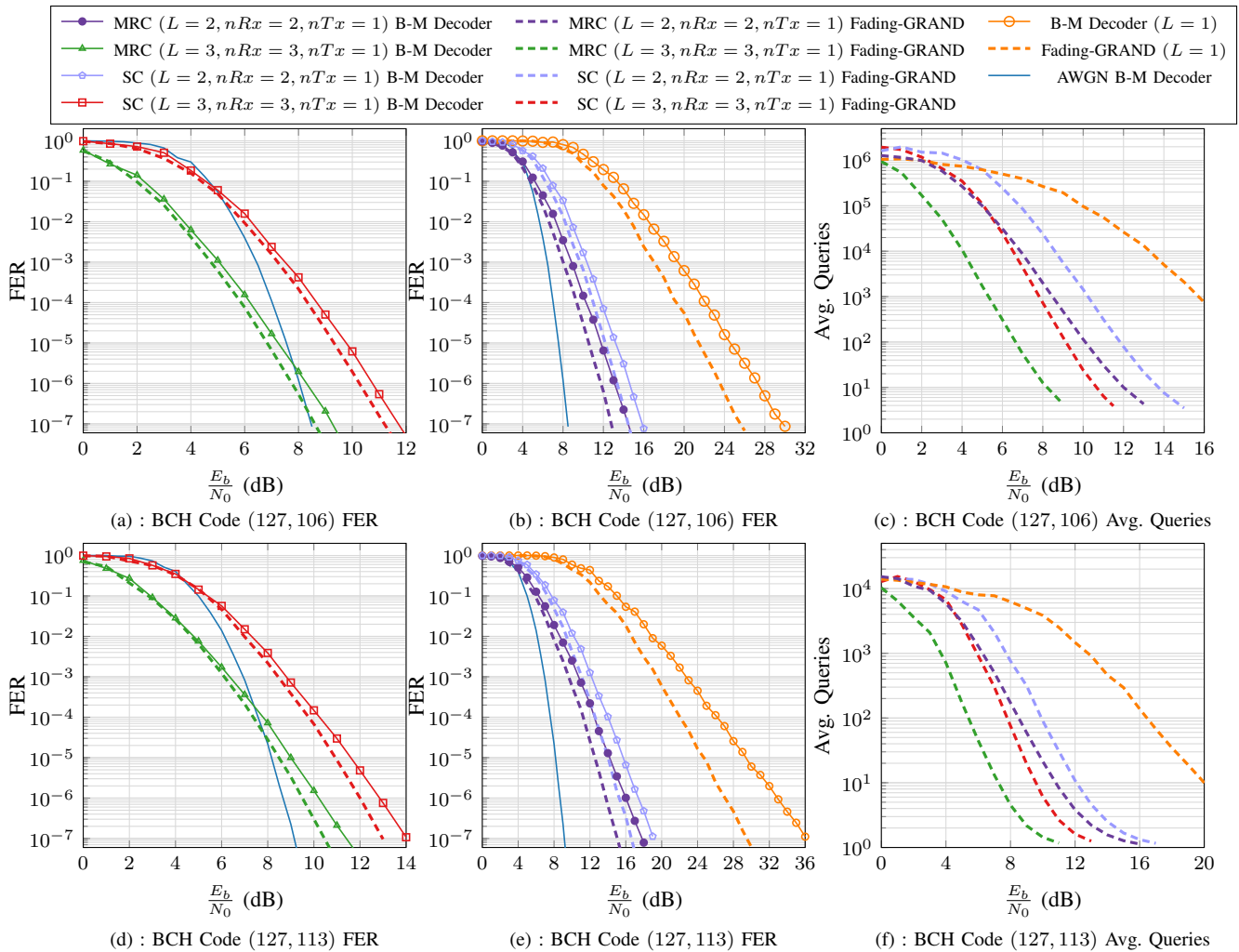
\begin{figure*}[t]
  \centering
  \begin{tikzpicture}[spy using outlines = {rectangle, magnification=2.0, connect spies}]
    \begin{groupplot}[group style={group name=fer_queries, group size= 3 by 2, horizontal sep=31pt, vertical sep=45pt},
      footnotesize,
      width=.7\columnwidth, height=.67\columnwidth, 
      xlabel=$\frac{E_b}{N_0}$ (dB),
      xmin=0, xmax=36, xtick={0,2,...,34},
      ymode=log,
      tick align=inside,
      grid=both, grid style={gray!30},
      /pgfplots/table/ignore chars={|},
      ]

\nextgroupplot[ylabel= FER, ytick pos=left, y label style={at={(axis description cs:-0.155,.5)},anchor=south},ymin=6e-8, ymax = 2,xmin=0, xmax=12]
      \addplot[mark=none , Paired-1, semithick]  table[x=Eb/N0, y=FER] {data_fading/BCH/BCH_127_106/BCH_127_106_BM.txt};\label{gp:plot1_a}
      

      \addplot[mark=triangle,mark options={scale=0.7} , Paired-3, semithick]  table[x=SNR, y=FER] {data_fading/BCH/BCH_127_106/FER_Fading_MRC_Div_3.txt};\label{gp:plot1_n}
      \addplot[mark=square,mark options={scale=0.7} , Paired-5, semithick]  table[x=SNR, y=FER] {data_fading/BCH/BCH_127_106/FER_Fading_SelComb_Div_3.txt};\label{gp:plot1_o}
      \addplot[mark=none , Paired-3, densely dashed, very thick]  table[x=SNR, y=FER] {data_fading/BCH/BCH_127_106/FER_Fading_MRC_3_SRGrand_bestParameter.txt};\label{gp:plot1_p}
      \addplot[mark=none , Paired-5, densely dashed, very thick]  table[x=SNR, y=FER] {data_fading/BCH/BCH_127_106/FER_Fading_Sel_3_SRGrand_bestParameter.txt};\label{gp:plot1_q}

      \coordinate (spypoint1) at (axis cs:14.3,1.0e-7);
      \coordinate (magnifyglass1) at (axis cs:11.5,0.35e-0);
      \coordinate (spypoint2) at (axis cs:11.4,1.0e-7);
      \coordinate (magnifyglass2) at (axis cs:3.0,0.3e-4);
      \coordinate (spypoint3) at (axis cs:9.0,1.0e-7);
      \coordinate (magnifyglass3) at (axis cs:3.0,0.3e-6);

      \coordinate (top) at (rel axis cs:0,1);

      \nextgroupplot[ylabel= FER, ytick pos=left, y label style={at={(axis description cs:-0.155,.5)},anchor=south},ymin=6e-8, ymax = 2, xmin=0, xmax=32, xtick={0,4,...,36},]
      \addplot[mark=none , Paired-1, semithick]  table[x=Eb/N0, y=FER] {data_fading/BCH/BCH_127_106/BCH_127_106_BM.txt};
      
      \addplot[mark=o , Paired-7, semithick]  table[x=Eb/N0, y=FER] {data_fading/BCH/BCH_127_106/BCH_127_106_BM_Rayleigh.txt};\label{gp:plot1_d}
      \addplot[mark=none , Paired-7, densely dashed, very thick]  table[x=SNR, y=FER] {data_fading/BCH/BCH_127_106/FER_Fading_SRGrand_bestParameter_linear.txt};\label{gp:plot1_i}

        \addplot[mark=*,mark options={scale=0.7} , Paired-9, semithick]  table[x=SNR, y=FER] {data_fading/BCH/BCH_127_106/FER_Fading_MRC_Div_2.txt};\label{gp:plot1_e}
      \addplot[mark=pentagon,mark options={scale=0.7} , Paired-14, semithick]  table[x=SNR, y=FER] {data_fading/BCH/BCH_127_106/FER_Fading_SelComb_Div_2.txt};\label{gp:plot1_g}
      \addplot[mark=none , Paired-9, densely dashed, very thick]  table[x=SNR, y=FER] {data_fading/BCH/BCH_127_106/FER_Fading_MRC_SRGrand_bestParameter.txt};\label{gp:plot1_j}
      \addplot[mark=none , Paired-14, densely dashed, very thick]  table[x=SNR, y=FER] {data_fading/BCH/BCH_127_106/FER_Fading_Sel_SRGrand_bestParameter.txt};\label{gp:plot1_l}
      
      \nextgroupplot[ylabel=Avg. Queries, ytick pos=left,y label style={at={(axis description cs:-0.12,.5)},anchor=south},ymin=1, ymax = 5e6, xmin=0, xmax=16,xtick={0,2,...,16},]
      \addplot[mark=none , Paired-3, densely dashed, very thick]  table[x=SNR, y=Comp] {data_fading/BCH/BCH_127_106/FER_Fading_MRC_3_SRGrand_bestParameter.txt};

      \addplot[mark=none , Paired-5, densely dashed, very thick]  table[x=SNR, y=Comp] {data_fading/BCH/BCH_127_106/FER_Fading_Sel_3_SRGrand_bestParameter.txt};

      \addplot[mark=none , Paired-14, densely dashed, very thick]  table[x=SNR, y=Comp] {data_fading/BCH/BCH_127_106/FER_Fading_Sel_SRGrand_bestParameter.txt};

      \addplot[mark=none , Paired-7, densely dashed, very thick]  table[x=SNR, y=Comp] {data_fading/BCH/BCH_127_106/FER_Fading_SRGrand_bestParameter_linear.txt};

      \addplot[mark=none , Paired-9, densely dashed, very thick]  table[x=SNR, y=Comp] {data_fading/BCH/BCH_127_106/FER_Fading_MRC_SRGrand_bestParameter.txt};

\nextgroupplot[ylabel= FER, ytick pos=left, y label style={at={(axis description cs:-0.155,.5)},anchor=south},ymin=6e-8, ymax = 2,xmin=0, xmax=14]
      \addplot[mark=none , Paired-1, semithick]  table[x=Eb/N0, y=FER] {data_fading/BCH/BCH_127_113/BCH_127_113_BM.txt};

      \addplot[mark=triangle,mark options={scale=0.7} , Paired-3, semithick]  table[x=SNR, y=FER] {data_fading/BCH/BCH_127_113/FER_Fading_MRC_Div_3.txt};
      \addplot[mark=square,mark options={scale=0.7} , Paired-5, semithick]  table[x=SNR, y=FER] {data_fading/BCH/BCH_127_113/FER_Fading_SelComb_Div_3.txt};

      \addplot[mark=none , Paired-3, densely dashed, very thick]  table[x=SNR, y=FER] {data_fading/BCH/BCH_127_113/FER_Fading_MRC_3_SRGrand_bestParameter.txt};
      \addplot[mark=none , Paired-5, densely dashed, very thick]  table[x=SNR, y=FER] {data_fading/BCH/BCH_127_113/FER_Fading_Sel_3_SRGrand_bestParameter.txt};

      \coordinate (spypoint4) at (axis cs:17,1.0e-7);
      \coordinate (magnifyglass4) at (axis cs:14,0.35e-0);

      \coordinate (top) at (rel axis cs:0,1);

      \nextgroupplot[ylabel= FER, ytick pos=left, y label style={at={(axis description cs:-0.155,.5)},anchor=south},ymin=6e-8, ymax = 2, xmin=0, xmax=36, xtick={0,4,...,36},]
      \addplot[mark=none , Paired-1, semithick]  table[x=Eb/N0, y=FER] {data_fading/BCH/BCH_127_113/BCH_127_113_BM.txt};

      \addplot[mark=o,mark options={scale=0.7} , Paired-7, semithick]  table[x=Eb/N0, y=FER] {data_fading/BCH/BCH_127_113/BCH_127_113_BM_Rayleigh.txt};
      \addplot[mark=none , Paired-7, densely dashed, very thick]  table[x=SNR, y=FER] {data_fading/BCH/BCH_127_113/FER_Fading_SRGrand_bestParameter_linear.txt};

      \addplot[mark=*,mark options={scale=0.7} , Paired-9, semithick]  table[x=SNR, y=FER] {data_fading/BCH/BCH_127_113/FER_Fading_MRC_Div_2.txt};
      \addplot[mark=pentagon,mark options={scale=0.7} , Paired-14, semithick]  table[x=SNR, y=FER] {data_fading/BCH/BCH_127_113/FER_Fading_SelComb_Div_2.txt};

      \addplot[mark=none , Paired-9, densely dashed, very thick]  table[x=SNR, y=FER] {data_fading/BCH/BCH_127_113/FER_Fading_MRC_SRGrand_bestParameter.txt};

      \addplot[mark=none , Paired-14, densely dashed, very thick]  table[x=SNR, y=FER] {data_fading/BCH/BCH_127_113/FER_Fading_Sel_SRGrand_bestParameter.txt};

      \nextgroupplot[ylabel=Avg. Queries, ytick pos=left,y label style={at={(axis description cs:-0.12,.5)},anchor=south},ymin=1, ymax = 5e4, xmin=0, xmax=20,xtick={0,4,...,20},]

      \addplot[mark=none , Paired-3, densely dashed, very thick]  table[x=SNR, y=Comp] {data_fading/BCH/BCH_127_113/FER_Fading_MRC_3_SRGrand_bestParameter.txt};

      \addplot[mark=none , Paired-5, densely dashed, very thick]  table[x=SNR, y=Comp] {data_fading/BCH/BCH_127_113/FER_Fading_Sel_3_SRGrand_bestParameter.txt};

      \addplot[mark=none , Paired-14, densely dashed, very thick]  table[x=SNR, y=Comp] {data_fading/BCH/BCH_127_113/FER_Fading_Sel_SRGrand_bestParameter.txt};

      \addplot[mark=none , Paired-7, densely dashed, very thick]  table[x=SNR, y=Comp] {data_fading/BCH/BCH_127_113/FER_Fading_SRGrand_bestParameter_linear.txt};

      \addplot[mark=none , Paired-9, densely dashed, very thick]  table[x=SNR, y=Comp] {data_fading/BCH/BCH_127_113/FER_Fading_MRC_SRGrand_bestParameter.txt};

      \coordinate (bot) at (rel axis cs:1,0);
    \end{groupplot}
    \node[below = 1cm of fer_queries c1r1.south] {\footnotesize (a) : BCH Code $(127,106)$ FER};
    \node[below = 1cm of fer_queries c2r1.south] {\footnotesize (b) : BCH Code $(127,106)$ FER};
    \node[below = 1cm of fer_queries c3r1.south] {\footnotesize (c) : BCH Code $(127,106)$ Avg. Queries};
    \node[below = 7cm of fer_queries c1r1.south] {\footnotesize (d) : BCH Code $(127,113)$ FER};
    \node[below = 7cm of fer_queries c2r1.south] {\footnotesize (e) : BCH Code $(127,113)$ FER};
    \node[below = 7cm of fer_queries c3r1.south] {\footnotesize (f) : BCH Code $(127,113)$ Avg. Queries};
    \path (top|-current bounding box.north) -- coordinate(legendpos) (bot|-current bounding box.north);
    \matrix[
    matrix of nodes,
    anchor=south,
    draw,
    inner sep=0.2em,
    draw
    ]at(legendpos)
    { 
      \ref{gp:plot1_e}& \scriptsize MRC $(L = 2, nRx=2, nTx=1)$  B-M Decoder &[5pt]
      \ref{gp:plot1_j}& \scriptsize MRC $(L = 2, nRx=2, nTx=1)$ Fading-GRAND &[5pt]
      \ref{gp:plot1_d}& \scriptsize B-M Decoder $(L=1)$ \\
      \ref{gp:plot1_n}& \scriptsize MRC $(L = 3, nRx=3, nTx=1)$ B-M Decoder &[5pt]
      \ref{gp:plot1_p}& \scriptsize MRC $(L = 3, nRx=3, nTx=1)$ Fading-GRAND &[5pt]
      \ref{gp:plot1_i}& \scriptsize Fading-GRAND $(L=1)$ \\
      \ref{gp:plot1_g}& \scriptsize SC $(L = 2, nRx=2, nTx=1)$ B-M Decoder &[5pt]
      \ref{gp:plot1_l}& \scriptsize SC $(L = 2, nRx=2, nTx=1)$ Fading-GRAND &[5pt]
      \ref{gp:plot1_a}& \scriptsize AWGN B-M Decoder  \\
      \ref{gp:plot1_o}& \scriptsize SC $(L = 3, nRx=3, nTx=1)$  B-M Decoder &[5pt]
      \ref{gp:plot1_q}& \scriptsize SC $(L = 3, nRx=3, nTx=1)$  Fading-GRAND \\
       }; 
  \end{tikzpicture}
   \vspace*{-1em}
  \caption{\label{fig:fer_bch_127_106} Comparing Fading-GRAND and BCH Berlekamp-Massey (B-M) decoding of BCH codes.}
   \vspace*{-1em}
\end{figure*}


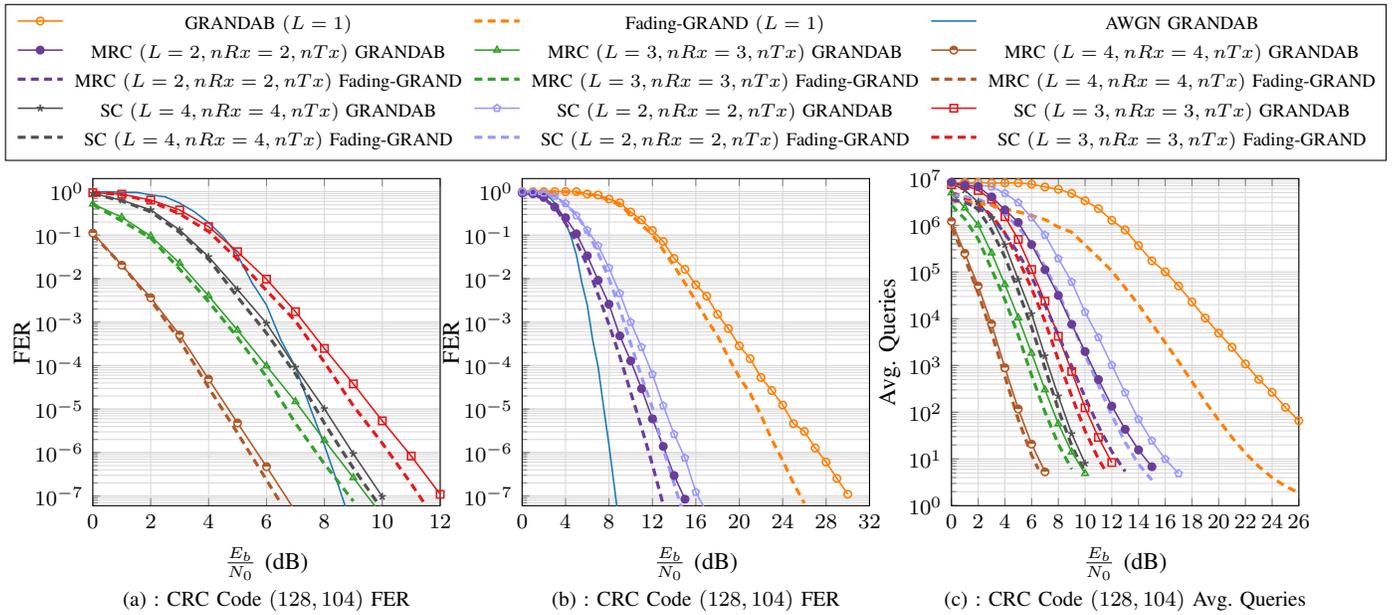
\begin{figure*}[t]
  \centering
  \begin{tikzpicture}[spy using outlines = {rectangle, magnification=2.0, connect spies}]
    \begin{groupplot}[group style={group name=fer_queries, group size= 3 by 2, horizontal sep=31pt, vertical sep=45pt},
      footnotesize,
      width=.7\columnwidth, height=.67\columnwidth, 
      xlabel=$\frac{E_b}{N_0}$ (dB),
      xmin=0, xmax=36, xtick={0,2,...,34},
      ymode=log,
      tick align=inside,
      grid=both, grid style={gray!30},
      /pgfplots/table/ignore chars={|},
      ]

\nextgroupplot[ylabel= FER, ytick pos=left, y label style={at={(axis description cs:-0.155,.5)},anchor=south},ymin=6e-8, ymax = 2,xmin=0, xmax=12]
      \addplot[mark=none , Paired-1, semithick]  table[x=Eb/N0, y=FER] {data_fading/CRC/128_104/AWGN.txt};\label{gp:plot_crc_a}
      
      \addplot[mark=halfcircle* ,mark options={scale=0.7}, Paired-11, semithick]  table[x=SNR, y=FER] {data_fading/CRC/128_104/FER_Fading_MRC_Div_4.txt};\label{gp:plot_crc_f}
      \addplot[mark=star,mark options={scale=0.7} , Paired-13, semithick]  table[x=SNR, y=FER] {data_fading/CRC/128_104/FER_Fading_SelComb_Div_4.txt};\label{gp:plot_crc_h}

      \addplot[mark=triangle,mark options={scale=0.7} , Paired-3, semithick]  table[x=SNR, y=FER] {data_fading/CRC/128_104/FER_Fading_MRC_Div_3.txt};\label{gp:plot_crc_n}
      \addplot[mark=square,mark options={scale=0.7} , Paired-5, semithick]  table[x=SNR, y=FER] {data_fading/CRC/128_104/FER_Fading_SelComb_Div_3.txt};\label{gp:plot_crc_o}
      
      \addplot[mark=none , Paired-11, densely dashed, very thick]  table[x=SNR, y=FER] {data_fading/CRC/128_104/FER_Fading_MRC_4_SRGrand_bestParameter.txt};\label{gp:plot_crc_ll}
      \addplot[mark=none , Paired-13, densely dashed, very thick]  table[x=SNR, y=FER] {data_fading/CRC/128_104/FER_Fading_Sel_4_SRGrand_bestParameter.txt};\label{gp:plot_crc_m}

      \addplot[mark=none , Paired-3, densely dashed, very thick]  table[x=SNR, y=FER] {data_fading/CRC/128_104/FER_Fading_MRC_3_SRGrand_bestParameter.txt};\label{gp:plot_crc_p}
      \addplot[mark=none , Paired-5, densely dashed, very thick]  table[x=SNR, y=FER] {data_fading/CRC/128_104/FER_Fading_Sel_3_SRGrand_bestParameter.txt};\label{gp:plot_crc_q}

      \coordinate (spypoint1) at (axis cs:14.3,1.0e-7);
      \coordinate (magnifyglass1) at (axis cs:11.5,0.35e-0);
      \coordinate (spypoint2) at (axis cs:11.4,1.0e-7);
      \coordinate (magnifyglass2) at (axis cs:3.0,0.3e-4);
      \coordinate (spypoint3) at (axis cs:9.0,1.0e-7);
      \coordinate (magnifyglass3) at (axis cs:3.0,0.3e-6);

      \coordinate (top) at (rel axis cs:0,1);

      \nextgroupplot[ylabel= FER, ytick pos=left, y label style={at={(axis description cs:-0.155,.5)},anchor=south},ymin=6e-8, ymax = 2, xmin=0, xmax=32, xtick={0,4,...,36},]
      \addplot[mark=none , Paired-1, semithick]  table[x=Eb/N0, y=FER] {data_fading/CRC/128_104/AWGN.txt};
      
      \addplot[mark=o,mark options={scale=0.7} , Paired-7, semithick]  table[x=SNR, y=FER] {data_fading/CRC/128_104/FER_Fading_CRC_128_104.txt};\label{gp:plot_crc_d}
      \addplot[mark=none , Paired-7, densely dashed, very thick]  table[x=SNR, y=FER] {data_fading/CRC/128_104/FER_Fading_SRGrand_bestParameter.txt};\label{gp:plot_crc_i}

       \addplot[mark=*,mark options={scale=0.7} , Paired-9, semithick]  table[x=SNR, y=FER] {data_fading/CRC/128_104/FER_Fading_MRC_Div_2.txt};\label{gp:plot_crc_e}
      \addplot[mark=pentagon,mark options={scale=0.7} , Paired-14, semithick]  table[x=SNR, y=FER] {data_fading/CRC/128_104/FER_Fading_SelComb_Div_2.txt};\label{gp:plot_crc_g}
      \addplot[mark=none , Paired-9, densely dashed, very thick]  table[x=SNR, y=FER] {data_fading/CRC/128_104/FER_Fading_MRC_SRGrand_bestParameter.txt};\label{gp:plot_crc_j}
      \addplot[mark=none , Paired-14, densely dashed, very thick]  table[x=SNR, y=FER] {data_fading/CRC/128_104/FER_Fading_Sel_SRGrand_bestParameter.txt};\label{gp:plot_crc_l}

      \nextgroupplot[ylabel=Avg. Queries, ytick pos=left,y label style={at={(axis description cs:-0.12,.5)},anchor=south},ymin=1, ymax = 1e7, xmin=0, xmax=26,xtick={0,2,...,26},]

      \addplot[mark=halfcircle*,mark options={scale=0.7} , Paired-11, semithick]  table[x=SNR, y=Comp] {data_fading/CRC/128_104/FER_Fading_MRC_Div_4.txt};
      \addplot[mark=none , Paired-11, densely dashed, very thick]  table[x=SNR, y=Comp] {data_fading/CRC/128_104/FER_Fading_MRC_4_SRGrand_bestParameter.txt};
      \addplot[mark=star,mark options={scale=0.7} , Paired-13, semithick]  table[x=SNR, y=Comp] {data_fading/CRC/128_104/FER_Fading_SelComb_Div_4.txt};
      \addplot[mark=none , Paired-13, densely dashed, very thick]  table[x=SNR, y=Comp] {data_fading/CRC/128_104/FER_Fading_Sel_4_SRGrand_bestParameter.txt};
      \addplot[mark=triangle,mark options={scale=0.7} , Paired-3, semithick]  table[x=SNR, y=Comp] {data_fading/CRC/128_104/FER_Fading_MRC_Div_3.txt};
      \addplot[mark=none , Paired-3, densely dashed, very thick]  table[x=SNR, y=Comp] {data_fading/CRC/128_104/FER_Fading_MRC_3_SRGrand_bestParameter.txt};

      \addplot[mark=square,mark options={scale=0.7} , Paired-5, semithick]  table[x=SNR, y=Comp] {data_fading/CRC/128_104/FER_Fading_SelComb_Div_3.txt};
      \addplot[mark=none , Paired-5, densely dashed, very thick]  table[x=SNR, y=Comp] {data_fading/CRC/128_104/FER_Fading_Sel_3_SRGrand_bestParameter.txt};

      \addplot[mark=pentagon,mark options={scale=0.7} , Paired-14, semithick]  table[x=SNR, y=Comp] {data_fading/CRC/128_104/FER_Fading_SelComb_Div_2.txt};
      \addplot[mark=none , Paired-14, densely dashed, very thick]  table[x=SNR, y=Comp] {data_fading/CRC/128_104/FER_Fading_Sel_SRGrand_bestParameter.txt};

      \addplot[mark=o,mark options={scale=0.7} , Paired-7, semithick]  table[x=SNR, y=Comp] {data_fading/CRC/128_104/FER_Fading_CRC_128_104.txt};
      \addplot[mark=none , Paired-7, densely dashed, very thick]  table[x=SNR, y=Comp] {data_fading/CRC/128_104/FER_Fading_SRGrand_bestParameter.txt};

      \addplot[mark=*,mark options={scale=0.7} , Paired-9, semithick]  table[x=SNR, y=Comp] {data_fading/CRC/128_104/FER_Fading_MRC_Div_2.txt};
      \addplot[mark=none , Paired-9, densely dashed, very thick]  table[x=SNR, y=Comp] {data_fading/CRC/128_104/FER_Fading_MRC_SRGrand_bestParameter.txt};

      \coordinate (bot) at (rel axis cs:1,0);
    \end{groupplot}
    \node[below = 1cm of fer_queries c1r1.south] {\footnotesize (a) : CRC Code $(128,104)$ FER};
    \node[below = 1cm of fer_queries c2r1.south] {\footnotesize (b) : CRC Code $(128,104)$ FER};
    \node[below = 1cm of fer_queries c3r1.south] {\footnotesize (c) : CRC Code $(128,104)$ Avg. Queries};
    \path (top|-current bounding box.north) -- coordinate(legendpos) (bot|-current bounding box.north);
    \matrix[
    matrix of nodes,
    anchor=south,
    draw,
    inner sep=0.2em,
    draw
    ]at(legendpos)
    { \ref{gp:plot_crc_d}& \scriptsize  GRANDAB $(L=1)$  &[0.1pt]
      \ref{gp:plot_crc_i}& \scriptsize  Fading-GRAND $(L=1)$ &[0.1pt]
      \ref{gp:plot_crc_a}& \scriptsize AWGN GRANDAB   \\
      \ref{gp:plot_crc_e}& \scriptsize MRC $(L=2,nRx=2,nTx)$  GRANDAB  &[0.1pt]
      \ref{gp:plot_crc_n}& \scriptsize MRC $(L=3,nRx=3,nTx)$ GRANDAB  &[0.1pt]
      \ref{gp:plot_crc_f}& \scriptsize MRC $(L=4,nRx=4,nTx)$ GRANDAB  \\ 
      \ref{gp:plot_crc_j}& \scriptsize MRC $(L=2,nRx=2,nTx)$ Fading-GRAND &[0.1pt]
      \ref{gp:plot_crc_p}& \scriptsize MRC $(L=3,nRx=3,nTx)$ Fading-GRAND &[0.1pt]
      \ref{gp:plot_crc_ll}& \scriptsize MRC $(L=4,nRx=4,nTx)$ Fading-GRAND \\
      \ref{gp:plot_crc_h}& \scriptsize SC $(L=4,nRx=4,nTx)$ GRANDAB &[0.1pt]
      \ref{gp:plot_crc_g}& \scriptsize SC $(L=2,nRx=2,nTx)$ GRANDAB  &[0.1pt]
      \ref{gp:plot_crc_o}& \scriptsize SC $(L=3,nRx=3,nTx)$  GRANDAB  \\
      \ref{gp:plot_crc_m}& \scriptsize SC $(L=4,nRx=4,nTx)$ Fading-GRAND &[0.1pt]
      \ref{gp:plot_crc_l}& \scriptsize SC $(L=2,nRx=2,nTx)$ Fading-GRAND &[0.1pt]
      \ref{gp:plot_crc_q}& \scriptsize SC $(L=3,nRx=3,nTx)$  Fading-GRAND \\
       }; 
  \end{tikzpicture}
  \vspace*{-2em}
  \caption{\label{fig:fer_crc_128_104} Comparison of decoding performance and average complexity of CRC code $(128,104)$ decoding via Fading-GRAND and GRANDAB decoder; For both GRANDAB and Fading-GRAND decoders, $nTx=1$.}
   \vspace*{-1em}
\end{figure*}

\section{Proposed Fading GRAND}

The steps of the proposed Fading GRAND are outlined in Algorithm \ref{alg:Fgrand}. The algorithm's inputs are the hard demodulated channel observation values ($\hat{\bm{y}}$) of size $n$, the maximum Hamming weight of the error patterns $(AB)$, the set of reliable indices $\mathbb{I}$ (explained in Section \ref{sec:TEP}), the parity check matrix $\bm{H}$ and the generator matrix $\bm{G}$. 

The hard-demodulated vector $\hat{\bm{y}}$ is subjected to a syndrome check (\ref{eq:constraint}) in the first phase of decoding (line 2) and if codebook membership criterion (\ref{eq:constraint}) is satisfied, decoding is presumed to be successful (line 4). Otherwise, Fading-GRAND generates TEPs ($\bm{e}$) for a particular Hamming Weight $HW$ $(HW\in [1,AB])$, such that the indices of the non-zero elements of generated error patterns ($\bm{e}$) do not belong to the reliable set $\mathbb{I}$ (line 7). The generated TEPs ($\bm{e}$) are then applied sequentially to $\hat{\bm{y}}$ and the resulting vector $\hat{\bm{y}}\oplus\bm{e}$ is then queried for codebook membership (line 8). If the codebook membership criterion (\ref{eq:constraint}) is met, then $\bm{e}$ represents the guessed noise and $\hat{\bm{c}} \triangleq \hat{\bm{y}}~\oplus~\bm{e}$ represents the estimated codeword from which the message ($\hat{\bm{u}}$) is recovered (line 9). Otherwise, error patterns for larger Hamming weights or the remaining error patterns for that Hamming weight are generated. When all error patterns for the maximum Hamming weight $(AB)$ have been generated and checked for codebook membership (\ref{eq:constraint}), the decoding procedure is terminated.

\subsection{Test Error Pattern (TEP) generation}\label{sec:TEP}

TEP generation, which is the key distinction between different GRAND variants, is at the center of GRAND decoding. GRANDAB generates TEPs in ascending Hamming weight order up to Hamming weight $AB$ ($HW_{max}=AB$). Combining a TEP with the received vector $\bm{\hat{y}}$ corresponds to flipping certain bits of that vector ($\bm{\hat{y}}$). The TEPs generated in ascending Hamming weight order for $n=6$ and $AB=4$ are shown in Fig. \ref{fig:TEP_GRANDAB} (a), where each column corresponds to a TEP and a dot corresponds to a flipped bit location of $\bm{\hat{y}}$. TEPs with Hamming weight $1$ are generated first, followed by TEPs with Hamming weight $2$,$3$ and $4$, as illustrated in Fig. \ref{fig:TEP_GRANDAB}. 

We propose a modification to the GRANDAB's TEP generation that reduces the fading impact in the Rayleigh fading communication channel. As shown in Algorithm \ref{alg:Fgrand}, the proposed Fading-GRAND is based on generating a set of reliable indices $\mathbb{I}$ that is used to generate only a subset of GRANDAB TEPs. Please note that, like GRANDAB, the Fading-GRAND generates TEPs in ascending Hamming weight order. The proposed Fading-GRAND TEP generation with $\mathbb{I}=\{3\}$ is shown in Fig. \ref{fig:TEP_GRANDAB} (b), where only the TEPs that exclude the indices in $\mathbb{I}$ are generated ($\mathds{1}_{\{\lambda_1,\lambda_2,\lambda_3,\lambda_4\}}\forall\{\lambda_1,\lambda_2,\lambda_3,\lambda_4\} \not\in \mathbb{I}$ where $\lambda_i\in[1,6]$ and $i\in[1,4]~(AB=4)$).

For a particular threshold ($\Delta$), the set $\mathbb{I}$  can be generated as 

\begin{equation}
i\in\mathbb{I} \iff w_i \geq \Delta, \forall~ i \in [1, n] ~\text{where}
\label{eq:threshold}
\end{equation}
\[ w_i = \begin{cases} \mbox{$\vert h_{\max}\vert$,} & \mbox{if } \text{SC}~(L>1) \\ \mbox{$\frac{\sum_{a=1}^{L}\vert h_{a}\vert}{L}$,} & \mbox{if } \text{MRC}~(L>1) \\ \mbox{$\vert h_{i}\vert$,} & \mbox{$L=1$} \end{cases} \]

For a Rayleigh fading channel with no spatial diversity branch $(L=1)$, Fig. \ref{fig:line_RLC_128_104} (a) illustrates the frame error rate (FER) performance for decoding RLC $(128,104)$ \cite{RLC2} with the proposed Fading-GRAND, with varying threshold ($\Delta$) values at different $\frac{E_b}{N_0}$. The optimal threshold $(\tilde{\Delta})$ is the threshold value that achieves the lowest FER at a constant $\frac{E_b}{N_0}$  $(\tilde{\Delta}=\argmin_{\Delta} \left (\mathrm{FER}(\frac{E_b}{N_0})\right)$. The optimal thresholds ($\tilde{\Delta}$) for each $\frac{E_b}{N_0}$ are plotted in Fig. \ref{fig:line_RLC_128_104} (b), and the equation of the best fitting line can be extracted from the optimal thresholds as $\tilde{\Delta}=m\times(\frac{E_b}{N_0}) + b$, where $m$ is the slope of the best-fit line and $b$ is the y-intercept. Please note that the optimal threshold ($\tilde{\Delta}$) for different $\frac{E_b}{N_0}$ can be computed offline and stored in a look-up table (LUT) for a linear $(n,k)$ block code.  As a result, using these pre-computed thresholds ($\tilde{\Delta}$) and the CSI from the receiver, the set $\mathbb{I}$ can be easily generated during Fading-GRAND decoding.

It should be noted that in \cite{SRGRAND} a thresholded variant of GRANDAB named Symbol Reliability GRAND (SRGRAND) for the AWGN channel was proposed. SRGRAND thresholds the soft input channel observation values ($\bm{y}$), classifies individual bits as reliable or unreliable, then applies GRANDAB solely to the unreliable bits. Fading-GRAND differs from SRGRAND in that it applies thresholding on the CSI (\ref{eq:threshold}) of the fading channel; hence, it requires the hard-demodulated vector ($\hat{\bm{y}}$) rather than soft input channel values ($\bm{y}$). Furthermore, the maximum Hamming Weight ($HW_{max}$) of the Fading-GRAND is equal to the $HW_{max}$ of the underlying GRANDAB ($HW_{max}=AB$), whereas $HW_{max}$ of the SRGRAND can be greater than the GRANDAB ($AB\leq{HW}_{max}\leq{n}$) under the maximum number of queries constraint. In addition to reducing complexity, restricting the $HW_{max}\leq{AB}$ leads to simpler hardware implementation, as shown in \cite{GRANDAB-VLSI}.

Figure \ref{fig:fer_RLC_128_104} compares the GRANDAB performance to the proposed Fading-GRAND\footnote{For both GRANDAB and Fading-GRAND decoders the $HW_{max}=4$ for BCH code $(127,106)$, CRC code $(128,104)$ and RLC $(128,104)$, whereas the $HW_{max}=3$ for BCH code $(127,113)$.} for decoding RLC $(128,104)$, which uses the optimal thresholds ($\tilde{\Delta}$) for generating the set $\mathbb{I}$, on Rayleigh fading channel. In addition, the GRANDAB decoding performance on the AWGN channel is included for reference. At a target FER of $10^{-7}$, the Fading-GRAND outperforms the GRANDAB by $8$ dB in decoding performance ($L=1$), as shown in Fig. \ref{fig:fer_RLC_128_104} (b). When the number of spatial diversity branches $(L)$  is increased to $2$, which corresponds to a scenario in which the transmitter is equipped with one antenna $(nTx = 1)$ and the receiver has two antennas $(nRx = 2)$, the proposed fading GRAND results in a $\sim3$ dB improvement over GRANDAB for both SC and MRC combining techniques as shown in Fig. \ref{fig:fer_RLC_128_104} (a).

\subsection{Computational complexity}

The computational complexity of GRAND and its variants can be represented in terms of the number of codebook membership queries required. In addition, the complexity can be divided into two categories: worst-case complexity, which corresponds to the maximum number of codebook membership queries required, and average complexity, which corresponds to the average number of codebook membership queries required. It is worth noting that as channel conditions improve, the average complexity of GRAND and its variants decreases rapidly, because transmissions subject to light noise are quickly decoded \cite{Duffy19TIT}. 

For a code length of $n$, the worst-case number of queries for the both Fading-GRAND and GRANDAB decoder is $\sum\limits_{i=1}^{AB} \binom{n}{i}$ \cite{Duffy19TIT}, where $AB$ is the maximum Hamming weight of the TEPs generated. Figure \ref{fig:fer_RLC_128_104} (c) compares the average complexity for Fading-GRAND and GRANDAB decoder for decoding RLC $(128,104)$. With $L=1$ and at $\frac{E_b}{N_0} =26$ dB, the GRANDAB decoder requires an average of $70$ queries whereas Fading-GRAND requires an average of $1.5$ queries at the same $\frac{E_b}{N_0}=26$ dB point (corresponding to a target FER of $10^{-7}$). With $L=2$ and at $\frac{E_b}{N_0}$ = $14$ dB, MRC with GRANDAB requires $15$ queries on average, whereas Fading-GRAND requires $2.6$ queries at $\frac{E_b}{N_0}$ = $14$ (corresponding to target FER of $10^{-7}$). As demonstrated in Fig. \ref{fig:fer_RLC_128_104} (c) at $E_b/N_0$ = $15$ dB with $L=2$, SC with GRANDAB requires $22$ queries, while Fading-GRAND requires $3.3$ queries at the same $\frac{E_b}{N_0}$ point corresponding to a target FER of $10^{-7}$. As a result, the proposed complexity of Fading-GRAND is $\sim \frac{1}{46}\times$ the complexity of GRANDAB for $L=1$ and $ \frac{1}{6}\times \sim \frac{1}{5}\times$ the complexity of GRANDAB for $L=2$, as shown in Fig. \ref{fig:fer_RLC_128_104} (c).

\section{Performance Evaluation}

In this section, we evaluate the proposed Fading-GRAND in terms of decoding performance and computational complexity for distinct classes of channel codes. For each class of channel codes, Table \ref{table:tableParam} presents the parameters $(m,b)$ for the equation of the best fitting line for optimal thresholds ($\tilde{\Delta}$). In the previous section, we discussed the framework for computing these thresholds at various $\frac{E_b}{N_0}$ and approximating them with a best-fitting line. For the numerical simulation results presented in this work, the proposed Fading-GRAND uses the equation $\tilde{\Delta}=m\times(\frac{E_b}{N_0}) + b$ with parameters indicated in Table \ref{table:tableParam} to threshold the CSI and generate set $\mathbb{I}$.

Figure \ref{fig:fer_bch_127_106} compares the FER performance of Fading-GRAND with the traditional Berlekamp-Massey (B-M) decoder \cite{Berlekamp68,Massey69} for decoding BCH code $(127,106)$ and BCH code $(127,113)$ on Rayleigh fading channel. In addition, for reference, the B-M decoding results \cite{AFF3CT} for the AWGN channel are included. For decoding BCH code $(127,106)$, the proposed Fading-GRAND outperforms the B-M decoder by $4$ dB for $L = 1$ (Fig. \ref{fig:fer_bch_127_106} (a)), $1.5\sim2$ dB for $L= 2$, and $\sim0.5$ dB for $L= 3$ at a target FER of $10^{-7}$ (Fig. \ref{fig:fer_bch_127_106} (b)). Similarly, for decoding BCH code $(127,113)$, the proposed Fading-GRAND outperforms the B-M decoder by $6.5$ dB, $2.5\sim3$ dB, and $1$ dB for $L=1,2,\text{and}~3$ respectively at target FER of $10^{-7}$ as shown in figures \ref{fig:fer_bch_127_106} (d) and (e). Furthermore, the figures \ref{fig:fer_bch_127_106} (c) and (f) illustrate the average computation complexity of Fading-GRAND for decoding BCH code $(127,106)$ and BCH code $(127,113)$. As demonstrated in Fig. \ref{fig:fer_bch_127_106} (c) and (f), the average complexity of the Fading-GRAND decreases significantly as channel conditions improve. 

Figure \ref{fig:fer_crc_128_104} compares the decoding performance  as well as the average computational complexity of Fading-GRAND and GRANDAB with CRC codes \cite{Peterson61}. 
For CRC code $(128,104)$ the generator polynomial is \texttt{0xB2B117}. Fading-GRAND outperforms GRANDAB by $4.5$ dB for $L=1$, $2\sim2.5$ dB for $L=2$, and $0.7$ dB for $L=3$ at a target FER of $10^{-7}$. 
The average complexity of Fading-GRAND is $\frac{1}{34}\times$ the complexity complexity of GRANDAB for $L=1$ ($\frac{E_b}{N_0} =26$ dB). Whereas for $L=2$ ($\frac{E_b}{N_0} =13$ dB for MRC and $\frac{E_b}{N_0} =14$ dB for SC) and for $L=3$ ($\frac{E_b}{N_0} =9$ dB for MRC and $\frac{E_b}{N_0} =11.5$ dB for SC) the average complexity of Fading-GRAND is $\frac{1}{7}\times\sim\frac{1}{8}\times$ and $\frac{1}{2}\times$ the complexity of GRANDAB respectively.
However, for $L = 4$, the Fading-GRAND results in a $0.2$ dB improvement and almost similar complexity to the GRANDAB. 

\begin{table}[t]
\centering
\caption{\label{table:tableParam} Parameters for Fading-GRAND ($\tilde{\Delta}=m\times\frac{E_b}{N_0} + b$)  for decoding linear $(n,k)$ block codes.}
\vspace*{-0.6em}
\begin{adjustbox}{max width=\columnwidth}
\begin{tabular}{ccccc}
\toprule
$(n,k)$                          &                                                                           & L & m        & b     \\
\cmidrule(l){1-1}\cmidrule(l){3-3}\cmidrule(l){4-4}\cmidrule(l){5-5}
\cmidrule(l){1-5}
\multirow{4}{*}{RLC $(128,104)$} & Rayleigh                                                                  & 1 & -0.0376   &  1.228  \\
\cmidrule(l){2-2}
                               & MRC                                                                      & 2 & -0.0579      & 1.325 \\
\cmidrule(l){2-2}
                               & SC                                                                        & 2 & -0.04975     & 1.614    \\
\cmidrule(l){1-5}
\multirow{7}{*}{BCH $(127,106)$} & Rayleigh                                                                  & 1 & -0.02944 & 1.002 \\
\cmidrule(l){2-2}
                               & \multirow{2}{*}{MRC}                                                     & 2 & -0.04833   & 1.191      \\
                               &                                                                              & 3 & -0.0378   & 1.275    \\
\cmidrule(l){2-2}
                               & \multirow{2}{*}{SC}                                                       & 2 & -0.04238    & 1.563    \\
                               &                                                                               & 3 & -0.05249   & 1.836    \\
\cmidrule(l){1-5}
\multirow{7}{*}{BCH $(127,113)$} & Rayleigh                                                                  & 1 &  -0.02165  &  0.7924   \\
\cmidrule(l){2-2}
                               & \multirow{2}{*}{MRC}                                                      & 2 & -0.04044 & 0.9037  \\
                               &                                                                           & 3 & -0.0588  & 1.174      \\
\cmidrule(l){2-2}
                               & \multirow{2}{*}{SC}                                                       & 2 & -0.03669   & 1.244      \\
                               &                                                                           & 3 & -0.03385   &  1.404     \\
\cmidrule(l){1-5}
\multirow{7}{*}{CRC $(128,104)$} & Rayleigh                                                                  & 1 & -0.03467  & 1.222      \\
\cmidrule(l){2-2}
                               & \multirow{3}{*}{MRC}                                                      & 2 & -0.06626     &  1.47  \\
                               &                                                                           & 3 & -0.0541    & 1.541      \\
                               &                                                                           &  4 & -0.07166    & 1.734  \\
\cmidrule(l){2-2}
                               & \multirow{3}{*}{SC}                                                       & 2 & -0.04426     &  1.568 \\
                               &                                                                           & 3 & -0.03443   &  1.737     \\
                               &                                                                           & 4 & -0.02029     &  1.839 \\
\bottomrule
\end{tabular}
\end{adjustbox}
\vspace*{-1em}
\end{table}


\section{Conclusion and Future Works}
In this paper, we introduced Fading-GRAND, a hard-input variation of GRAND designed for multipath flat Rayleigh fading communication channels. Fading can significantly degrade the performance of wireless communication systems, hence diversity approaches are used to combat it. The proposed Fading-GRAND adapts its TEPs to the fading conditions of the underlying channel, outperforming a conventional channel code decoder in both scenarios where spatial diversity is present and when there is no spatial diversity. For decoding BCH code $(127,106)$ and BCH code $(127,113)$, numerical simulation results show that the Fading-GRAND outperforms the B-M decoder by $0.5\sim6.5$dB  at a target FER of $10^{-7}$. Similarly, when compared to the previously proposed GRANDAB, Fading-GRAND outperforms GRANDAB by $0.2\sim8$dB for decoding CRC $(128,104)$ code and RLC $(128,104)$. Additionally, the average complexity of Fading-GRAND, at $\frac{E_b}{N_0}$ corresponding to target FER of $10^{-7}$, is $\frac{1}{2}\times\sim \frac{1}{46}\times$ the complexity of GRANDAB. This enhanced decoding performance and low complexity of Fading-GRAND makes it appealing for applications with stringent performance and complexity constraints. The Fading-GRAND ushers GRAND research into practical multipath fading channel scenarios, and it can be further investigated for other spatial and frequency diversity techniques (MIMO and OFDM) as well as other fading channels (Rician and Nakagami).

\balance
\bibliographystyle{IEEEtran}
\bibliography{IEEEabrv, FadingGRAND}
\end{document}